\documentclass{PoS}
\usepackage{xcolor}
\usepackage{lineno}
\usepackage[utf8]{inputenc}

\title{Redshift of the Blazar KUV 00311-1938: Modeling the EBL Absorption}

\ShortTitle{Redshift of the Blazar KUV 00311-1938}

\author{M. Fernandez Alonso\\
         Department of Physics, Pennsylvania State University, University Park, PA, USA\\
        E-mail: \email{mkf5479@psu.edu}}

\author{A. Pichel\\
        Instituto de Astronomía y Física del Espacio (IAFE, UBA-CONICET), Buenos Aires, Argentina\\}

\author{\speaker{A.C. Rovero}\\
        Instituto de Astronomía y Física del Espacio (IAFE, UBA-CONICET), Buenos Aires, Argentina\\}

\abstract{AGNs are believed to be the source of very energetic cosmic rays which, in turn, produce gamma rays throughout the interactions with matter. Therefore, the study of their very-high-energy (VHE) gamma-ray spectra is fundamental to understand the cosmic-ray acceleration mechanisms in these objects. Blazars are the most common type of AGNs populating gamma-ray catalogs, among these sources BL-Lacs present featureless or very weak emission/absorption lines spectra. Given that the extragalactic VHE gamma-ray catalog is poorly populated (about 70 blazars), having a redshift estimation of every single source is important for the study of cosmic rays origin. In this work we use EBL absorption as a tool to estimate KUV 00311-1938 redshift by comparing HE gamma-ray observations with modeled EBL absorbed flux points. KUV 00311-1938 is a very interesting BL-Lac; a priori it can be one of the most distant blazars since it has an estimated redshift in the range 0.5 to 1.54. Assuming that there is no spectral cutoff, we consider the intrinsic VHE spectrum to be the extrapolation of the Fermi-LAT spectrum and model the intergalactic absorption by using the main current EBL models. Comparing these results with the VHE measured point by H.E.S.S., we give an estimate for the redshift of the blazar within observation errors.}

\FullConference{36th International Cosmic Ray Conference -ICRC2019-\\
		July 24th - August 1st, 2019\\
		Madison, WI, U.S.A.}

\begin{document}

\section{EBL Absorption and KUV 00311-1938} 
\label{sec:abs}

\begin{figure}[]
  \centering
  \includegraphics[width=.8\textwidth]{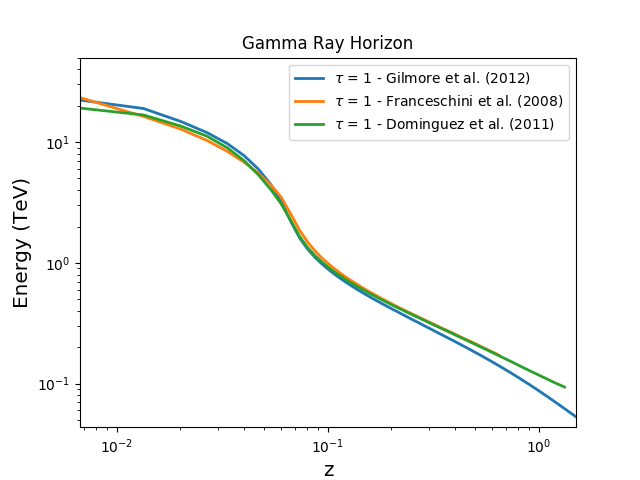}
  \caption{Gamma ray horizon for different EBL models.}
  \label{fig:GammaHorizon}
\end{figure}

The extragalactic background light (EBL) is the diffuse cosmological light covering the range from UV to far-IR, composed by the integrated contribution of radiative and re-emission processes throughout the history of the Universe. 
The Universe is opaque for gamma rays in the very-high-energy (VHE: E$>$100 GeV) range. Photon absorption in the intergalactic (IG)  photon backgrounds is energy dependent and starts to become substantial at TeV energies \cite{1966PhRvL..16..252G}. In particular, VHE gamma rays from jets of AGN can interact with photons in the IR-UV range present in the EBL and photons from the Cosmic Microwave Background (CMB), producing electron-positron pairs and defining a gamma-ray horizon. Given an EBL model, the opacity ($\tau$) depends on both the energy of the photon and its redshift (z). Figure \ref{fig:GammaHorizon} shows possible gamma-ray horizons ($\tau$=1) for different EBL models.

Multi-TeV gamma rays are expected to be highly attenuated for 0.1 < z < 1 and almost completely attenuated for z > 1. The EBL spectral distribution is still uncertain, and so is the intrinsic spectra of distant blazars, for which the absorption is considerable. However, certain physical motivated assumptions can be made about them. In the GeV range, absorption is minimal, therefore it it reasonable to expect the observed spectrum of a distant source to be similar to the intrinsic one for this range of energies \cite{2011ApJ...733...77O}. A basic assumption consists on considering the intrinsic spectrum in the TeV range as a prolongation of the spectrum in the GeV regime. It is important to mention that this assumption neglects the fact that part of the observed photons in the high-energy (HE: E>100 MeV) regime are the result of electromagnetic cascades generated by TeV gamma rays in the intergalactic medium. To estimate the overall effect of this component in the observed HE spectrum requires simulations and further assumptions, for example the EBL density model. This is indeed an interesting study on itself but it escapes the scope of our analysis.

Blazars are a subclass of active galactic nuclei (AGN), and the most numerous HE and VHE gamma-ray emitters. Their spectra are often dominated by non-thermal radiation. Most blazars are BL-Lac objects, which have a featureless or very weak emission/absorption lines spectra. This implies that the determination of their redshifts is extremely difficult to achieve; as of today only about 75\% of VHE gamma-ray blazars have a measured spectroscopic redshift. 

KUV 0031-1938 is a BL-Lac blazar with unknown redshift, detected by Fermi-LAT at HE \cite{2009ApJ...700..597A,2011ApJ...743..171A} and by H.E.S.S. at VHE in 2012 \cite{2012AIPC.1505..490B}. 
It was also detected previously in the X-ray band by ROSAT \cite{2000ApJS..129..547B,2000AN....321....1S}.
Several attempts to determine its redshift were made with no success, or contradictions among them. In \cite{2007A&A...470..787P} they observed the object with the ESO 3.6 m telescope for 900 s in July 2001, estimating a tentative redshift of z=0.610 based in the optical spectrum with very weak lines. The object was also observed using the 6dF fibre-fed multi-object spectrograph at the United Kingdom Schmidt Telescope (UKST), but despite the fact that this instrument provides a better $S/N$ ratio, spectral lines were not found \cite{2009MNRAS.399..683J}. Another study claimed to have found the MgII doublet, estimating a lower limit of z=0.50507 for this object \cite{2012AIPC.1505..566P}, and this was later confirmed in \cite{2013AJ....146..127S}. Observations using the X spectrograph on the VLT from UV to NIR and did not find the absorption lines \cite{2014A&A...565A..12P}, estimating an upper limit of z=1.54.
All these results place KUV 0031-1938 in the very interesting condition of potentially being one of the most distant blazars detected at VHE, with redshift in a range poorly populated by other VHE blazars, i.e. 0.506$<$z$<$1.54.

In this analysis we use Fermi-LAT observations of the blazar KUV 00311-1938 to derive an intrinsic TeV spectrum. We then simulate the absorption process for different models and redshifts to compare the results with the available HESS data to provide a redshift estimation for the source.

\section{Fermi Analysis}
The GeV spectrum of KUV 00311-1938 (4FGL J0033.5-1921 in the 4FGL catalog) was obtained using 2 years of Fermi-LAT \texttt{PASS8} data from December 2009 to December 2011, corresponding with the same period of time in which HESS observed the same source. The LAT analysis was performed with the instrument response function \texttt{P8R2\_SOURCE\_V6} using a binned maximum-likelihood method implemented in the ScienceTools: gtlike. Data is binned in 8 energy bins per decade covering the range 100 MeV - 800 GeV, and the chosen spectral model is a log-parabola (eq. \ref{eq:logP}), which is in better agreement with this source data according to the 4FGL catalog \cite{2019arXiv190210045T}.
\begin{equation}
\label{eq:logP}
\frac{dN}{dE}=N_0\left(\frac{E}{E_b}\right)^{-\left(\alpha+\beta log\left(E/E_b\right)\right)}
\end{equation}

Figure \ref{fig:FermiPlot} shows the resulting fitted log-parabola spectrum and flux points for this analysis. The parameters resulting from the fit are the following:

\begin{figure}[]
  \centering
  \includegraphics[width=1\textwidth]{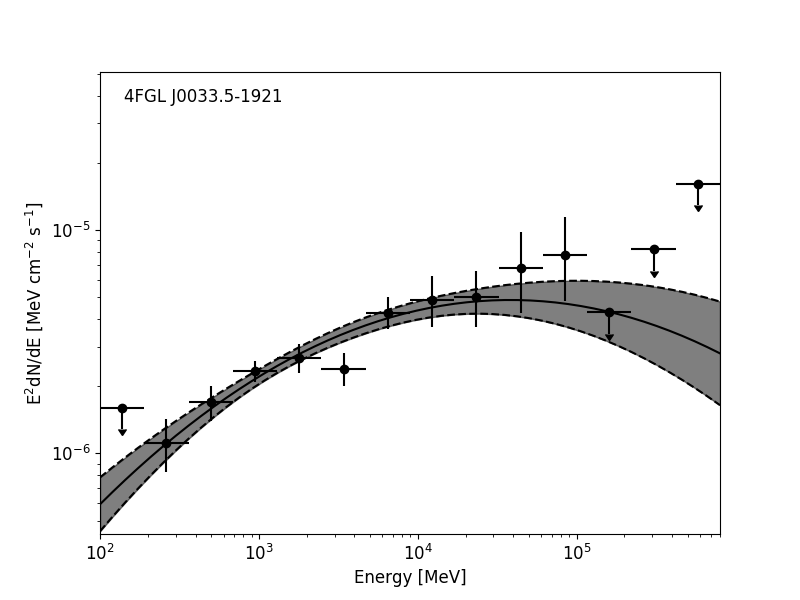}
  \caption{KUV00311-1938 Fermi-LAT flux points with fitted log-parabola function and corresponding 1$\sigma$ confidence band (grey).}
  \label{fig:FermiPlot}
\end{figure}

\begin{itemize}
\item $N_0$ = 1.15$\times$10$^{-12}\pm$ 8.55$\times$10$^{-14}$ MeV$^{-1}$ cm$^{-2}$ s$^{-1}$
\item $E_b$= 1504 MeV
\item $\alpha$= 1.615 $\pm$ 0.072
\item $\beta$= 0.06 $\pm$ 0.02
\end{itemize}

The fit parameter results are in good agreement with those published in Fermi-LAT latest catalog \cite{2019arXiv190210045T} for this source.

\section{Redshift estimation}
\begin{figure}[]
  \centering
  \includegraphics[width=1\textwidth]{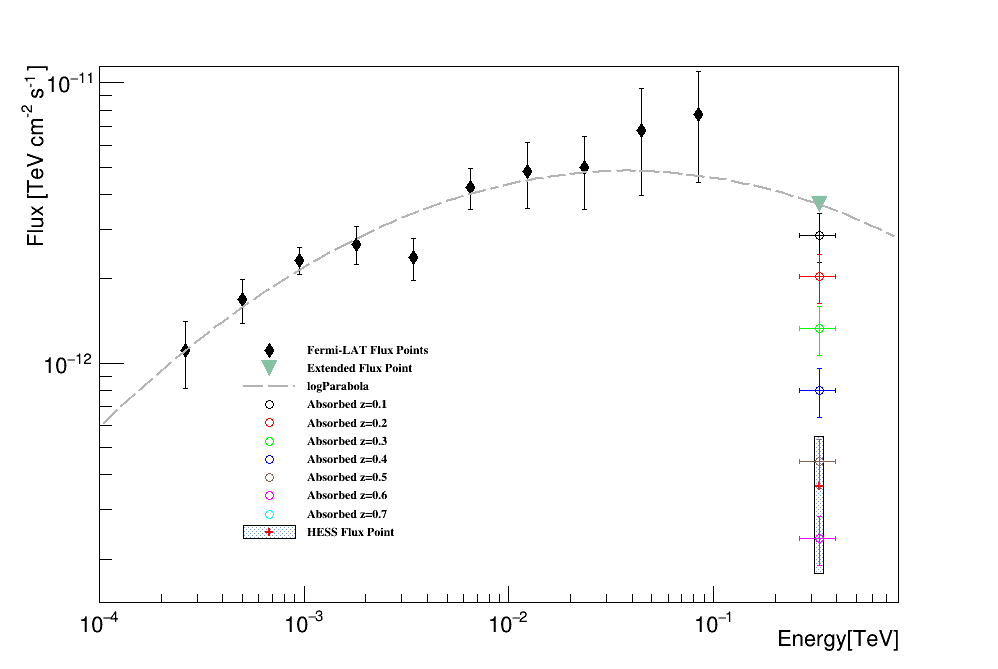}
  \caption{Prolonged log-parabola model extracted from Fermi-LAT analysis (light grey - dashed) and absorbed flux points (color circles) at E$_{dec}$ for different redshifts and assuming the EBL model proposed in \cite{Dominguez}. HESS flux point is displayed as a box delimited by the flux and energy errors.}
  \label{fig:SpectrumComplete}
\end{figure}

HESS has derived a gamma-ray spectrum for KUV 00311-1938 using a forward folding techniche \cite{2012AIPC.1505..490B}. The differential flux at the decorrelation energy (E$_{dec}$ = 0.33 TeV) is $\phi_{E_{dec}}$ = (3.33 $\pm$ 1.02 stat $\pm$ 0.67 sys) $\times$ 10$^{-12}$ cm$^{-2}$ s$^{-1}$ TeV$^{-1}$. We chose this point as the observed flux point to compare with different absorbed flux points. 
The intrinsic TeV flux of the source at $E_{dec}$ was obtained from the prolongation of the log-parabola given by equation \ref{eq:logP} and following the assumption explained in section \ref{sec:abs}. The absorbed flux points are obtained through the following equation:
\begin{equation}
 \centering
\left( \frac{dN}{dE} \right)_{abs}= \left( \frac{dN}{dE} \right)_{int} e^{-\tau\left( E,z\right) }
\label{eq_abs}
\end{equation}
where $\left( \frac{dN}{dE} \right)_{int}$ is the intrinsic spectrum flux point taken from the prolongation of the Fermi-LAT spectrum, $\left( \frac{dN}{dE} \right)_{abs}$ is the absorbed spectrum flux point, and $\tau\left(E,z\right) $ is the optical depth for a photon of energy E and redshift z. The absorbed flux points are calculated for different redshifts and assuming different EBL models proposed in \cite{Franceschini2008A&A...487..837F}, \cite{Gilmore} and \cite{Dominguez}. Each flux point was assigned a 25\% error in line with average IACT responses \cite{2011ApJ...733...77O}. The 1 sigma contour plot (butterfly plot) of the Fermi spectrum fit was used to explore the effect of slope variations in the resultant VHE observed flux points. No significant changes were observed in the VHE flux points for variations of the intrinsic spectral index within 1 sigma.
Figure \ref{fig:SpectrumComplete} shows the resulting absorbed flux points for different redshifts, the HESS flux point along with the Fermi-LAT observed flux points and the prolonged log-parabola model and assuming the EBL model proposed in \cite{Dominguez}.

The agreement between HESS observed flux and the absorbed flux is calculated using the following equation:
\begin{equation}
\Delta F = \frac{\vert F_{abs} - F_{HESS}\vert}{\sqrt{\sigma^2_{F_{HESS}}+\sigma^2_{F_{abs}}}}
\end{equation}
where $F_{abs}$ - $\sigma_{F_{abs}}$ and $F_{HESS}$ - $\sigma_{F_{HESS}}$  correspond to the absorbed and HESS flux points - flux errors respectively.
Table \ref{tab:deltaF} shows the resulting $\Delta F$ values for different redshifts and EBL models. HESS measured flux point is consistent within 1$\sigma$ errors with the absorbed flux from a source at z=0.5 - z=0.6, depending on the EBL model. All three EBL models present similar results, which is expected given the similar gamma-ray horizons (see figure \ref{fig:GammaHorizon}). 
\begin{table}[]
\begin{center}
\begin{tabular}{c c c c }
\hline \hline
  Redshift & Gilmore & Franceschini & Dominguez\\
 \hline
0.1 & 4.10 & 4.16 & 4.15\\
0.2 & 3.56 & 3.74 & 3.72\\
0.3 & 2.55 & 2.98 & 2.94\\
0.4 & 1.07 & 1.79 & 1.71\\
0.5 & \colorbox{pink}{0.33} & \colorbox{pink}{0.41} & \colorbox{pink}{0.33}\\
0.6 & 1.21 & \colorbox{pink}{0.66} & \colorbox{pink}{0.73}\\
0.7 & 1.64 & 1.30 & 1.35\\
\hline \hline
\end{tabular}
\caption{Calculated $\Delta$F values for different redshift and for each EBL model. Values below $\Delta$F < 1 are highlighted.}
\label{tab:deltaF}
\end{center}
\end{table}

\section{Discussion}

Optical methods for estimating distances are often challenging, specially for objects like blazars and distant galaxies, where spectral features are not very well distinguished. The study of EBL absorption proves to be a useful tool in these cases, either to estimate or/and to obtain independent redshift constrains. The method used here is presented as a first approach to estimate the redshift of this blazar using EBL absorption as a tool. By comparing Fermi-LAT and HESS observations from similar time periods and under the assumption that the TeV intrinsic spectrum is a prolongation of the GeV spectrum, we found the measured flux to be consistent within 1$\sigma$ errors with the absorbed flux of a source at z=0.5 and z=0.6 depending on the EBL model. 
In the near future, more VHE observations of this blazar will probably give better information about its spectral properties and will allow a more complete study. Also, the detector response and typical statistical fluctuations must be properly taken into account to obtain adequate flux errors for the absorbed flux points. 

As we better understand EBL features and enhance gamma-ray instruments to observe distant blazars, indirect distance estimation methods similar to the one presented in this work will probably become widely used within the community.


\bibliographystyle{JHEP}

\bibliography{biblio_fernandez}



\end{document}